# Critical Analysis of Middleware Architectures for Large Scale Distributed Systems

Florin Pop*, Ciprian Dobre*, Alexandru Costan*, Mugurel Ionut Andreica*, Eliana Tirsa*,
Corina Stratan*, Valentin Cristea*

*Computer Science Department, Faculty of Automatic Control and Computer Science,
University Politehnica of Bucharest
(e-mails: {florin.pop, ciprian.dobre, alexandru.costan, mugurel.andreica, eliana.tirsa, corina.stratan,
valentin.cristea}@cs.pub.ro)

**Abstract:** Distributed computing is increasingly being viewed as the next phase of Large Scale Distributed Systems (LSDSs). However, the vision of large scale resource sharing is not yet a reality in many areas – Grid computing is an evolving area of computing, where standards and technology are still being developed to enable this new paradigm. Hence, in this paper we analyze the current development of middleware tools for LSDS, from multiple perspectives: architecture, applications and market research. For each perspective we are interested in relevant technologies used in undergoing projects, existing products or services and useful design issues. In the end, based on this approach, we draw some conclusions regarding the future research directions in this area.

## 1. INTRODUCTION

Distributed Systems have become very useful, especially in the case of scientific applications, where the processing of very large data volumes is necessary in a very short amount of time, as well as the storage of this data. Taking into account the tremendous popularity of complex distributed systems, favored by the rapid development of computing systems, the high speed networks and the Internet, it is clear that it is imperative, in order to achieve performances as high as possible in the utilization of these systems, to pick an optimal structure and architecture, but also scheduling and data replications algorithms for the distributed systems. This thing is particularly difficult, and even impossible, to be done by somebody without the help of a specialized program, because the prediction of the functioning of a distributed system without the aid of the mentioned program is only approximate and there may appear functioning errors in that distributed system.

In the world of distributed computing, Grid computing has emerged as an important new field, distinguished from conventional distributed computing by its focus on large-scale resource sharing, innovative applications, and, in some cases, high-performance orientation. Grids are semantically different from other distributed systems and therefore performance analysis through simulation techniques requires careful reconsideration.

The concept of Grid appeared in the 1990's and is best defined by Ian Foster, one of its initiators, as coordinated resource sharing and problem solving in dynamic, multi-institutional Virtual Organizations (VOs). Grid computing has gained an increasing importance since the 1990's, especially in the academic environments, offering the possibility to rapidly solve complex scientific problems.

Nevertheless, in the last years Grid computing has also begun to gain ground in the commercial environments, with the aid of some important investments made by the world's leading IT companies. As the investments in Grid technologies are expected to increase dramatically and the complexity of the computing resources is evolving, the research in this field will be a subject of interest for the computer science community in the next years.

Different examples of distributed systems are: clusters (as a part in a Grid), Grids, P2P System, Web based Systems.

We shall start this paper by introducing some fundamental architecture concepts. Section 3 presents the achievements in the two main categories of specific software for LSDS: middleware (which consists of services that provide resource management, information registration and discovery, remote process management, monitoring etc.) and applications (developed on top of the middleware). Section 4 presents the critical analysis of presented middleware. In Section 5 we will present some remarks and open issues for LSDS middleware tools.

## 2. MIDDLEWARE ARCHITECTURES

An important aspect of the LSDS is the architecture that defines the system components, specifying the purpose and function of these components, and indicates their interactions. LSDSs are based on one or a combination of two architectures: a protocol oriented architecture, and a service-oriented architecture.

Historically, the most important **Grid architecture** is that proposed by Ian Foster, Carl Kesselman and Steven Tuecke (members of the Globus Alliance) and described in the very well known paper "The Anatomy of the Grid" (Foster, 2001). The authors started from the idea that sharing resources asks for interoperability among potential participants in a VO. Since interoperability means protocols, the architecture they developed for Grids is a protocol architecture "with protocols defining the basic mechanisms by which VO users and resources negotiate, establish, manage, and exploit sharing relationships" (Foster, 2001). The open source reference implementation of key Grid protocols was the Globus Toolkit V2 (GT2) (Globus, 2009).

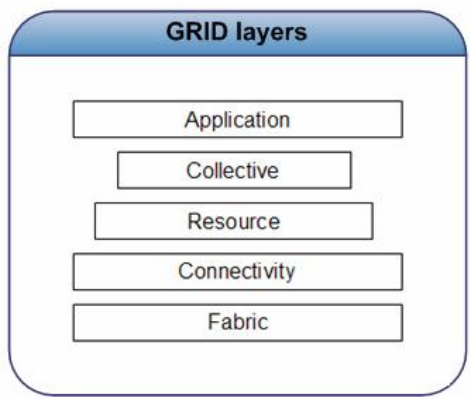

Fig. 1. Grid layered architecture (Foster, 2001).

This protocol architecture is a layered one (see Figure 1) and follows the "hourglass model", which requires the existence of a small set of protocols and abstractions, onto which many higher-level behaviors can be mapped, and which can themselves be mapped onto many underlying technologies (see the figure). The Fabric layer consists of resources (either physical or logical) to which the shared access is provided with the aid of the Grid protocols. Examples of such resources are: computational resources (like a computer cluster), storage systems (including distributed file systems), network resources, sensors. The Connectivity layer defines the base communication and authentication protocols required for Grid-specific network transactions. Communication protocols enable the exchange of data between Fabric layer resources and authentication protocols provide secure mechanisms for verifying the identity of users and resources. The Resource layer defines protocols (and APIs and SDKs) for the secure negotiation, initiation, monitoring, control, accounting, and payment of sharing operations on individual resources. The implementations of these protocols call Fabric layer functions to access and control local resources. At this level, only individual resources are regarded, not the global state of the distributed system. The Collective layer contains protocols and services that are not associated with specific resources, but with the interactions across collections of resources. The Collective components can implement a wide variety of sharing behaviors without placing new requirements on the resources being shared. The last layer, Application, contains the user applications that operate within a VO environment; the applications may themselves define protocols, services, and/or APIs and have a high degree of complexity ().

Later, this protocol-based architectural view of Grids has been augmented by the authors with a service-based view. In this view, the Grid is considered an "extensible set of services that respond to protocol messages" (Foster, 2002). Grid services may be aggregated to meet the requirements of Virtual Organizations. The new proposal, which has been termed Open Grid Services Architecture (OGSA), tries to align the Grid technologies with Web services technologies and to valorize the Web services properties that result from the use of the Web Service Definition language (WSDL) and from the separation of the neutral description of what the service offers to its contractors and the bindings corresponding to the actual service providers, namely: the automatic generation of client and server code from service descriptions in WSDL, service discovery, binding of service descriptions to interoperable network protocols, compatibility with emerging higher-level services (Foster, 2002). The test bed for OGSA was Globus toolkit 3, GT3.

OGSA describes standard mechanisms for creating, naming and discovering Grid services instances. In this model, computational resources, networks, storage resources, databases etc. are represented by services; a service can be defined as a network-enabled entity that provides some capability. A service-oriented view simplifies the virtualization (the encapsulation behind a common interface of diverse implementations) and addresses the need for standard interface definition mechanisms and local/remote transparency. In 2003, (Tuecke, 2003) provided a first specification of the OGSA concepts, called Open Grid Services Infrastructure (OGSI). The authors defined approaches for creating, naming and managing service instances, for declaring and inspecting service state data, for the notification of service state changes etc. The definitions are given as WSDL types, and can be used in combination in order to create complex Grid services. The architecture, adopted in the next version of the Globus Toolkit, GT4, is also service oriented (OGSA) but adopts a new paradigm for Grid services development, namely the Web Service Resource Framework (WSRF). The new approach is better focused on services than OGSA and provides a stronger compatibility with the existing Web Services tools.

***Peer-to-peer architecture*** is based on a network in which each node is considered having equivalent capabilities and responsibilities. P2P architecture classification is based on the network and application.

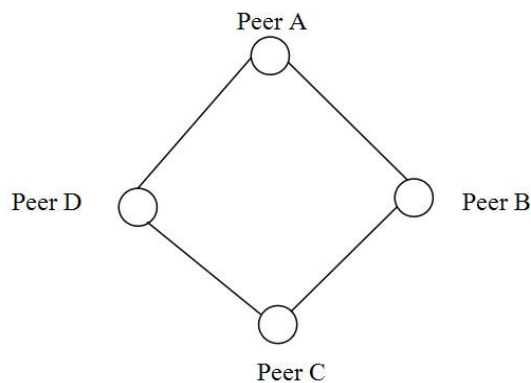

Fig. 2. Pure Peer Network (early Gnutella)

An example of a pure P2P file sharing network was the original design of Gnutella (released March 2000) in which the search function and content storage were totally decentralized, meaning that each function was conducted at the individual peer level. This design suffered from several technical weaknesses that have diminished its role as a competitive distribution platform (see Figure 2).

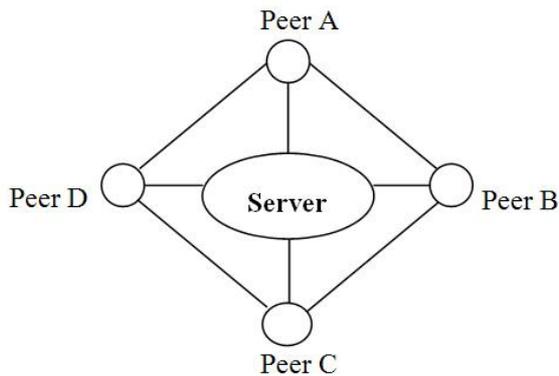

Fig. 3. Centrally Coordinated Peer Network (Napster)

One of the important aspects in P2P architecture is *Collaborative Distributed Computing*. It combines the idle or unused CPU processing power and/or free disk space of many computers in the network. Collaborative computing is most popular with science and biotech organizations where intense computer processing is required. The *Instant Messaging* allows users to send different types of messages in real-time. The *Affinity Communities* is the group of P2P networks that is based around file-sharing and became widely known and talked about due to the public legal issues surrounding the direct file sharing group, Napster (see Figure 3). Affinity Communities are based on users collaborating and searching other user's computers for information and files.

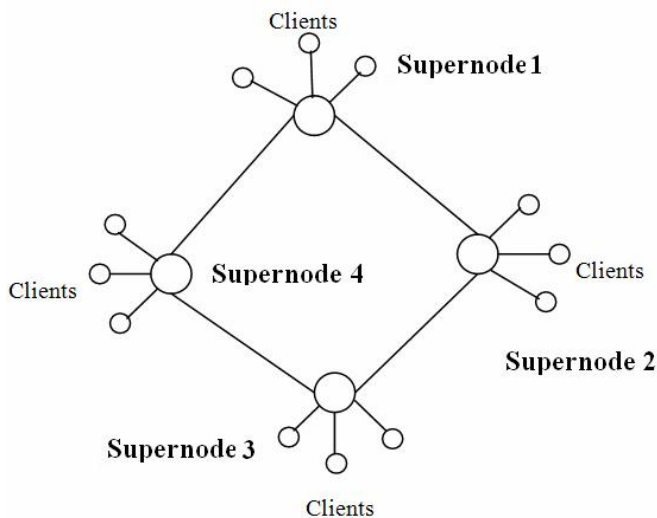

Fig. 4. Hierarchical Peer Network (Kazaa, Grokster)

The fault tolerant architecture for P2P systems is presented in Figure 4. The Hierarchical Peer Network considers different super-nodes which represents the communication point with clients.

P2P and Grid architectures differ from **Web-based systems' architecture** where some computers are dedicated to serving the others. In the modern approach, the web applications are based on services. An overview over standards-based web services shows that they differ in technology and in the applicability area. The success of the web services technology is conditioned by the existence of general open standards, available to any developer or user. The development of web services and applications must satisfy certain requirements: a web service must be able to answer to any client, regardless of the platform on which it is developed. A client must be able to retrieve the servers that can respond to its request through a web service.

The web service standards were defined to improve the interoperability and availability for users from different domains. Serving as a base for the development of Grid systems and applications, the Internet offers the support for the web services functionality. The diagram below, reproduced from (Tannenbaum 2002), describes the client-server model which is the base for web services design.

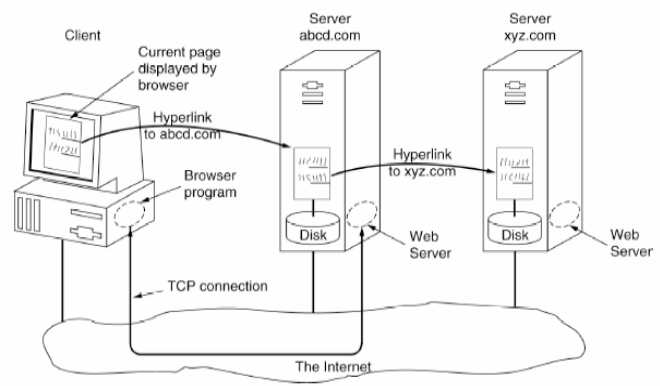

Fig. 5. Web-based system (Tanenbaum, 2002)

The communication between the service providers and the client's needs a common terminology; so that the exchanged information is understood by both of the parties in an effective manner (the XML standard offers the solution for this problem). *Simple Object Access Protocol* (SOAP) is a common protocol for representing the messages exchanged by web services (SOAP, 2009). The language which describes web services is *Web Services Description Language* (WSDL, 2009). *Universal Description, Discovery and Integration* (UDDI) defines the way in which the providers publish details about the services and the clients obtain the published information (UDDI, 2009).

The current requirements in developing Grid application impose the compliance with the standards described above. The applications that are currently designed for an architecture that includes support for web services are from various domains like economy (Buyya, 2002), industry and science (for example, simulations and data processing in nuclear physics and complex systems physics).

# 3. MIDDLEWARE TOOLS

Grid computing was based from the beginning on open standards and protocols, which were used from the first versions of dedicated software packages (e.g. Globus, Unicore). GT2 used, for example, the protocols from the TCP/IP stack for communication and authentication, but also developed new protocols, which took into account the particularities of network dynamics in Grid environments. Such a protocol is GridFTP, a high-performance, secure, reliable data transfer protocol based on FTP, optimized for high-bandwidth wide-area networks. Another Globus specific protocols are GRAM (Grid Resource Allocation and Management), for the remote submission of a computational request, GSI (Grid Security Infrastructure), GARA (Globus Architecture for Reservation and Allocation).

The next step in Grid technology was made by OGSI, which was centered on the concept of Grid Services. Grid Services are based on Web Services - technology that has the advantage of standardization, relying on Internet based standards like Simple Object Access Protocol (SOAP) and Web Services Description Language (WSDL). A Grid service is actually a Web service that conforms to a set of conventions and supports standard interfaces (Foster, 2002). The Grid services can maintain an internal state, which distinguishes an instance of the service from others. Most of the Grid software packages adopted the Web Services technology.

As the Web Services architecture evolved, the Web services community pointed out some shortcomings of OGSI, which are underlined in (Czajkowski, 2004): there are too many functions in a single specification, it does not work well with the existing Web services and XML tools (for example, JAX-RPC), it has too many similarities with the object oriented model (like the existence of instances and of an internal state), and it uses features from WSDL 2.0 which are not supported in WSDL 1.1.

These issues were addressed with the introduction of the WS-Resource framework, which aims to exploit the new Web services standards, especially WS-Addressing (standard that describes transport neutral mechanisms to address Web services). This refactoring has been done in three steps (Czajkowski, 2004):

- The introduction of the WS-Resource concept

- A better separation of function and exploitation of other Web services specifications

- A broader view of notification (for the state changes appearing in Web services)

In the WS-Resource Framework, the state is no longer stored within the service, but within the so-called resources. The composition of a stateful resource and a Web service is called a WSResource. WSRF and Grid services offer equivalent functionality, but WSRF has several advantages, like being easier to implement and exploit with the current Web services tools, and making a clearer separation between the service (which can be a simple message processor, both in WSRF and OGSI) and the resource, which stores the internal state.

The current state of the art in Grid architecture technologies is represented by the WSRF framework, which is being adopted in some of the most important Grid software toolkits and has reached sufficient maturity to be included in enterprise products (e.g., Univa Globus Enterprise).

*3.1. Resource Management*

Resource management in Grid implies a quite large number of functionalities, from resource discovery to scheduling, execution management, status monitoring and accounting. In this section, we shall focus on scheduling systems, and we shall present the monitoring functionalities and the Grid information systems in a further section. We shall introduce here some general issues, and then we shall present taxonomy of the scheduling systems and some details regarding the scheduling mechanisms used in the most important current Grid projects.

Wieder (2005) distinguishes between two cases of Grid systems with respect to their requirements on resource management capabilities: Case 1 is Specialized Grids for dedicated purposes, which are centered on a single or limited application domain and require high efficiency in execution. The Resource Management System (RMS) is adapted to the application, its workflow and the available resource description. Thus, the interfaces to the resources and the middleware are built according to the given requirements caused by the application scenario. While the Grid RMS is highly specialized, the handling for the user is often easier as the know-how of the application domain has been built into the system.

Case 2 is a Generic Grid Middleware, which has to cope with the complete set of the requirements to support applicability. Here, the Grid RMS is open for many different application scenarios. In comparison to the specialized Grids, generic interfaces that can be adapted to many frontend backend are required. However, the generic nature of this approach comes at the price of additionally overhead for providing information about the application. For instance, more information about a particular job has to be provided to the middleware, such as a workflow description, scheduling objectives, policies and constraints. The application knowledge cannot be built into the middleware, and therefore must be provided at the frontend level. In this case, the consideration of security requirements is an integral aspect, which is more difficult to solve. It is possible to hide the additional RMS complexity of generic Grid infrastructures from the users or their applications by specialized components, which might be built on top of a generic middleware. Nevertheless, it can be concluded that in general a generic Grid middleware will carry additional overhead with less efficiency at the expense of broader applicability.

Current research is mostly focusing on Case 1 in which solutions are built for a dedicated Grid scenario in mind. As mentioned before, these systems are usually more efficient and will therefore remain the favorite solution for many application domains. That is, Case 1 will not become obsolete if corresponding requirements and conditions exist. However, for creating future generation Grids suitable solutions are required for Case 2.

One of the most important components of a RMS is the scheduler, which distributes the applications on the Grid resources and usually also handles the execution management. We shall present as follows a brief taxonomy for scheduling systems.

*3.2. Data Management*

The Globus Toolkit provides several data management components, which can be classified in two categories: data movement services and data replication services.

For data movement, two main components are available: the GridFTP tools and the Reliable File Transfer (RFT) service.

The GridFTP protocol provides secure, robust, and fast data transfer. It is generally used for bulk data and it was defined by Global Grid Forum Recommendation GFD.020, RFC 959, RFC 2228 and RFC 2389. The GridFTP components are a powerful set of tools, but they have some weak points. For example, the client must have an open socket connection to the server during the transfer, who may not be interrupted; this makes the transfer of very large files difficult. The client is able to recover from remote failures (network outages, server failures, etc), but if the client itself or the client host fails, the recovery is not possible because the information needed for recovery is held in the client's memory.

These issues were solved in a new service included in the Globus Toolkit, called RFT (Reliable File Transfer). RFT is a Web Services Resource Framework (WSRF) compliant web service able to schedule intelligently the data movement operations. The user provides a list of source and destination URLs, and the service writes the information about the given transfer jobs into a database and starts moving the files. Once the service has taken your job request, interactions with it are similar to any job scheduler.

The Globus Toolkit provides two data replication services: the Replica Location Service (RLS), which is a basic data replication component, and the Data Replication Service (DRS), a higher-level service based on RLS and RFT (Reliable File Transfer).

*3.3. Systems Monitoring*

Operating a successful LSDS, network or computing facility requires vast amounts of monitoring information. Projects and organizations worldwide that need to track resource usage, network traffic, job distribution and many other quantities rely on monitoring systems to collect the information and present it in a way that allows them to make effective decisions. The systems also have to automatically troubleshoot and optimize very large grid and network systems.

While the initial target field of these applications were networks and Grid systems supporting data processing and analysis for global high energy and nuclear physics collaborations, monitoring tools are broadly applicable to many fields of data intensive science, and to the monitoring and management of major research and education networks.

An essential part of managing a global Data Grid is a monitoring system that is able to monitor and track the many site facilities, networks, and the many tasks in progress, in real time. The monitoring information gathered also is essential for developing the required higher level services, and components of the Grid system that provide decision support, and eventually some degree of automated decisions, to help maintain and optimize workflow through the Grid.

The relevant efforts invested in this domain are gathered in some major projects:

*GridICE* (GridICE 2009) is a distributed monitoring tool designed for Grid systems. It promotes the adoption of de-facto standard Grid Information Service interfaces, protocols and data models. Further, different aggregations and partitions of monitoring data are provided based on the specific needs of different user categories, each of them dealing with a different abstraction level of a Grid: the Virtual Organization level, the Grid Operation Center level, the Site Administration level and the End-User level. Being able to start from summary views and to drill down to details, it is possible to verify the composition of virtual pools or to sketch the sources of problems. A complete history of monitoring data is also maintained to deal with the need for retrospective analysis.

*R-GMA*. The Grid Monitoring Architecture (RGMA, 2009) consists of three components: Consumers, Producers and a directory service, (which we prefer to call a Registry). In the GMA Producers register themselves with the Registry and describe the type and structure of information they want to make available to the Grid. Consumers can query the Registry to find out what type of information is available and locate Producers that provide such information. Once this information is known the Consumer can contact the Producer directly to obtain the relevant data. The Registry communication is shown by a dotted line and the main flow of data by a solid line. The GMA architecture was devised for monitoring but it also makes an excellent basis for a combined information and monitoring system.

*Ganglia* (Ganglia, 2009) is a scalable distributed monitoring system for high-performance computing systems such as clusters and Grids. It is based on a hierarchical design targeted at federations of clusters. It leverages widely used technologies such as XML for data representation, XDR for compact, portable data transport, and RRDtool for data storage and visualization. It uses carefully engineered data structures and algorithms to achieve very low per-node overheads and high concurrency. The implementation is robust, has been ported to an extensive set of operating systems and processor architectures, and is currently in use on over 500 clusters around the world. It has been used to link clusters across university campuses and around the world and can scale to handle clusters with 2000 nodes.

The *MonALISA* (MonALISA, 2009) system is designed as an ensemble of autonomous multi-threaded, self-describing agent-based subsystems which are registered as dynamic services, and are able to collaborate and cooperate in performing a wide range of information gathering and processing tasks. These agents can analyze and process the information, in a distributed way, to provide optimization decisions in large scale distributed applications. An agent-based architecture provides the ability to invest the system with increasing degrees of intelligence, to reduce complexity and make global systems manageable in real time. The scalability of the system derives from the use of multithreaded execution engine to host a variety of loosely coupled self-describing dynamic services or agents and the ability of each service to register itself and then to be discovered and used by any other services, or clients that require such information. The system is designed to easily integrate existing monitoring tools and procedures and to provide this information in a dynamic, customized, self describing way to any other services or clients.

The scalability of the system derives from the use of a multi threaded engine to host a variety of loosely coupled self-describing dynamic services, the ability of each service to register itself and then to be discovered and used by any other services, or clients that require such information. The framework integrates many existing monitoring tools and procedures to collect parameters describing computational nodes, applications and network performance. Specialized mobile agents are used in the MonALISA framework to perform global optimization tasks or help and improve the operation of large distributed system by performing supervising tasks for different applications or real time parameters.

### 3.4. Security

Security Infrastructure has been motivated by the need of secure communication between entities over the Grid. It has to provide:

Authentication - allowing entities to interact knowing each other identities (on top of authenticated identities, authorization, logging and pricing schemes can be implemented).

- Privacy - guaranteeing protection of data exchanged both against tampering and unwanted access.
- Authorization - establishing and enforcing policies under which clients and services can interact.
- Delegation - enabling entities/resources to act on behalf of other entities/ resources/ clients.
- Single Sign-on - assuring that once an identity is authenticated/authorized, access can be obtained anywhere the entity is entitled to.

### 3.5. Applications

Applications can use registered services and tools (query, monitoring, discovery, factory, notification, security, registration, management, scheduling) along with grid infrastructure (Pearlman, 2003). We can define an application like a collection of work items or jobs that carry out a complex computing task by using grid resources. So, according with this definition, designing an application for grid computing is much easier if you know what to expect and which are the main work items. You should plan to use a development environment or toolkit specifically designed for grid applications, such as the Globus Toolkit and MonALISA or Ganglia.

Designing a grid application must consider three aspects:

- *Jobs*: flow, type of job, number of difficult jobs, depth of sub-jobs, redundant jobs execution, scavenging grid and job topology.
- *Data*: topology, data type – character sets and multimedia formats, amount of data, data separable per jobs, job data I/O, shared data access, temporary data space, time-sensitive data, data encryption.
- *Environment*: dependence of the OS, memory needed per job, compiler settings, library needed, runtime environment, application server, external application, hardware dependency, network bandwidth and scalability, security policy, single user interface, time constrains.

A running application in a grid system is called grid-enabled. For making an application grid-enabled, there are six strategies, according to (Kra, 2005). These strategies are:

- *Batch anywhere*. In this strategy, only the grid (not the application, the client, the user, or anything else) decides which node to use for the job.
- *Independent Concurrent Batch*. This supports multiple independent instances of the same application running concurrently.
- *Parallel Batch*. In this case, takes each user's batch work, subdivides it, disperses it out to multiple nodes, collects it, and then aggregates the results.
- *Service*. Service is follow-on to Independent Concurrent Batch, not follow-on to Parallel Batch. Service, it is not assumed that each client subdivides its work and spreads it over multiple service instances.
- *Parallel Service*. This strategy combines the service-oriented architecture of Service with the subdivided work model of Parallel Batch.
- *Tightly Coupled Parallel Programs*. This is the domain of specialized applications in engineering, physics, and biological modeling, such as finite state analysis.

It is important in this generation of LSDS applications to establish what type of strategies to use in the design process. For example, the run stage for your job must consider the first three strategies. The adapt process for job consider parallel batch, service and parallel service to be important and the last aspect exploit the cluster infrastructure considered the last one strategy.

## 4. CRITICAL ANALYSIS OF LSDS MIDDLEWARE

We present in the following table some characteristics for different type of LSDS middleware for Cluster, Grids, Web-based systems, Cloud Computing, P2P. The characteristics are presented from the applications side and include scope of the system, architecture, development models and technologies, supported platform.

|  | Cluster | Grid | Web-based system | Cloud Computing | P2P |
|---|---|---|---|---|---|
| **Scope** | High Performance Computing | Workflow execution | Client-Server Application | SOA Applications | File sharing Applications |
| **Architecture** | Centralized | Decentralized | Hierarchical | Hierarchical, Decentralized | Centralized, Hierarchical, Decentralized |
| **Development model** | Execution Job Object | Abstract Job Object | RPC based Object (RMI, Corba) | Web Services Object | Component Object |
| **Development technology** | Java, C/C++, Perl, Python | Java, C/C++, Perl, Python | Java, C/C++, C# | J2EE, .NET, WebSpere, Azure | Java, C++, C# |
| **Supported platform** | Unix, MacOS | Unix, MacOS | Unix, Windows, MacOS | Platform independent | Platform independent |
| **Users and applications** | SEE-GRID, EuroGrid, Grid Interoperability Project (GRIP) | EGEE, AppLeS, Ninf, Nimrod-G, NASA IPG, Gridbus Broker, eScience (UK), EU Data Grid. | Web 2.0 & 3.0 Applications (CSS, DHTML, JSP, Servlets, EJB, SaaS) | Identity (OAuth, OpenID), Integration (Amazon Simple Queue Service), Payments (Amazon Flexible Payments, Google Checkout, PayPal), Mapping (Google Maps, Yahoo! Maps), Search (Google Custom Search, Yahoo! BOSS), Others (Amazon Mechanical Turk) | Kazza, Napster, Gnutela |

## 5. CONCLUSIONS AND OPEN ISSUES

In this paper we present a critical analysis of middleware architectures for large scale distributed systems. We described the middleware architectures having different components with important role in the system's and application's life cycle. The open issues regarding the middleware tools are presented in the following.

In resource management, one of the most important research subjects is Service-Level Agreements (SLAs), with the aid of which the demands of the users and those of the resource owners can be better balanced. Some variants of SLAs have already been implemented within different research projects and they are about to be used on a wide scale.

Another aspect that will be considered in future LSDSs is the economic one – the users will be charged for the resources that they consume, and the schedulers will take this aspect into account, trying to obtain an optimal combination between the execution time for a job and the associated cost.

In the domain of information systems and monitoring, we observe as a main characteristic, the distribution of their architecture (in different forms, following each model – relational for RGMA, agent based and distributed services in MonALISA). This distribution implies, as a main direction in future research, the development of synchronization tools (for replicated data repositories containing monitoring information, for example), fault toleration, no single point of failure, prediction tools.

These developments have a broader range of applications, to the LSDS required for major experiments, and other data-intensive projects. The real time systems presented (such as MonALISA) also include much of the functionality required of the OGSA standardized services planned by the Global Grid Forum in the future.

Effective and robust integrated applications require higher level service components able to adapt to a wide range of requests, and changes in the state of the system (such as changes in the available resources, for example). These services should be capable of "learning" from previous experience, and apply "self-organizing neural network" or other heuristic algorithms to the information gathered, to optimize dynamically the system, by minimizing a set of "cost functions".

In what concerns LSDS security, the objectives concern seamless access for the clients entitled to use the addressed resources. The long time objective is to place in the Grid environment the ability to store, retrieve and manipulate client and service rights in order for the applications to meet autonomously authorization requirements. As we have seen, the authorization schemes are mainly identity based, being effective in small to middle size Grids. When it comes to interconnecting several Grids or accommodating a larger Grid, the mapping of identities to user accounts (as in the grid map file authorization scheme) does not scale. Nevertheless the identity of a user gives little or no information concerning user attributes, roles or associations with projects and institutions.

Current efforts are undergone in order to allow policy based access mechanisms, retrieval at interaction time of user attributes, rights and policies, and their combination to infer authorization decisions. Another area of research concerns fine grained authorization policy enforcement, advertisement and understanding, but, in the same time, provisioning of dynamic addressing and usage of Grid resources.

The new issues for LSDS resource are oriented to the development of WSRF-compliant architectures and the creation of new models that include Semantic Web technologies; in LSDS middleware, future research can be done in the domains of service-level agreements, market-based resource management, seamless and secure access for the users to LSDS resources. In what concerns the applications, most of the research efforts are directed to making as many scientific applications as possible "LSDS-enabled", so that they can benefit of the advantages of Grid technology.

## REFERENCES

Badia R. M., Olav Beckmann, Marian Bubak, Denis Caromel, Vladimir Getov, Ludovic Henrio, Stavros Isaiadis, Vladimir Lazarov, Maciek Malawski, Sofia Panagiotidi, Nikos Parlavantzas, Jeyarajan Thiyagalingam. *Lightweight Grid Platform: Design Methodology, CoreGRID Technical Report*, January 2006

Buyya R. *Economic-based Distributed Resource Management and Scheduling for Grid Computing*, Ph.D. Thesis, Monash University, Melbourne, Australia, 2002

Czajkowski K., Don Ferguson, Ian Foster, Jeff Frey, Steve Graham, Tom Maguire, David Snelling, Steve Tuecke. *From Open Grid Services Infrastructure to WSResource Framework: Refactoring & Evolution*, 2004.

Foster I., C. Kesselman, S. Tuecke (2001). The Anatomy of the Grid: Enabling Scalable Virtual Organizations. *International Journal of High Performance Computing Applications*, 15 (3) pp 200-222, 2001.

Foster I., C. Kesselman, J. Nick, S. Tuecke (2002). The Physiology of the Grid - An Open Grid Services Architecture for Distributed Systems Integration (extended version of Grid Services Architecture for *Distributed Systems Integration. IEEE Computer* 35(6), pp 37-46). Open Grid Service Infrastructure WG, Global Grid Forum. 2002

Globus Project - http://www.globus.org [Accessed 19th January 2009]

Ganglia project homepage - http://ganglia.sourceforge.net/ [Accessed 19th January 2009]

GridICE project - http://infnforge.cnaf.infn.it/gridice/ [Accessed 19th January 2009]

Homayounfar H., Wang F., Areibi S., (2002). An Advanced Peer-to-Peer Architecture using Autonomous Agents, *International Journal of Computers, Systems and Signals,* Vol.4 , No.2, 2002.

Kra D., *Six strategies for grid application enablement*. Retrieved from www.ibm.com on 18th January 2009.

Pearlman L., C. Kesselman, V. Welch, I. Foster, and S. Tuecke (2003). *The community authorization service: Status and future*. In Proceedings of the Conference for Computing in High Energy and Nuclear Physics, La Jolla, California, USA, Mar. 2003.

RGMA project homepage - http://www.r-gma.org/ [Accessed 19th January 2009]

SOAP: http://www.w3.org/TR/soap/, Accessed in January 2009

Tanenbaum A. S., van Steen M., (2002). *Distributed Systems*. Principles and paradigms, Prentice Hall 2002

Tuecke S., K. Czajkowski, I. Foster, J. Frey, S. Graham, C. Kesselman, T. Maquire, T. Sandholm, D. Snelling, P. Vanderbilt. *Open Grid services infrastructure* (OGSI) version 1.0, 2003.

UDDI: http://www.uddi.org/specification.html, Accessed in January 2009

Wieder P., U. Schwiegelsholn, R. Yahyapour. *Resource Management for Future Generation Grids. CEI, Dortmund, Germany*, 2005.

WSDL: http://www.w3.org/TR/wsdl, Accessed in January 2009